\documentclass[12pt,a4paper]{article}
\usepackage{amssymb}
\usepackage{graphicx,color}
\makeatletter
\def\rddots{\mathinner{\mkern1mu\raise\p@%
    \vbox{\kern7\p@\hbox{.}}\mkern2mu%
    \raise4\p@\hbox{.}\mkern2mu\raise7\p@\hbox{.}\mkern1mu}}
\makeatother
\makeatletter
\def\eqnarray{%
\stepcounter{equation}%
\let\@currentlabel=\theequation
\global\@eqnswtrue
\global\@eqcnt\z@
\tabskip\@centering
\let\\=\@eqncr
$$\halign to \displaywidth\bgroup\@eqnsel\hskip\@centering
$\displaystyle\tabskip\z@{##}$&\global\@eqcnt\@ne
\hfil$\displaystyle{{}##{}}$\hfil
&\global\@eqcnt\tw@$\displaystyle\tabskip\z@{##}$\hfil
\tabskip\@centering&\llap{##}\tabskip\z@\cr}
\makeatother
\newcommand{\ket}[1]{{\vert{#1}\rangle}}

\newcommand{\fukuso}{{\mathbf C}}
\newcommand{\real}{{\mathbf R}}
\newcommand{\futon}{{\bf N}}

\begin{document}

\title{\sl More on the Isomorphism $SU(2)\otimes SU(2)\cong SO(4)$}
\author{
  Kazuyuki FUJII
  \thanks{E-mail address : fujii@yokohama-cu.ac.jp }\ \ ,\ \ 
  Hiroshi OIKE
  \thanks{E-mail address : oike@tea.ocn.ne.jp }\quad and\ \ 
  Tatsuo SUZUKI
  \thanks{E-mail address : suzukita@gm.math.waseda.ac.jp }\\
  ${}^{*}$Department of Mathematical Sciences\\
  Yokohama City University\\
  Yokohama, 236--0027\\
  Japan\\
  ${}^{\dagger}$Takado\ 85--5,\ Yamagata, 990--2464\\
  Japan\\
  ${}^{\ddagger}$Department of Mathematical Sciences\\
  Waseda University\\
  Tokyo, 169--8555\\
  Japan\\
  }
\date{}
\maketitle
%
%
%
%
\begin{abstract}
  In this paper we revisit the isomorphism $SU(2)\otimes SU(2)\cong SO(4)$ 
  to apply to some subjects in Quantum Computation and Mathematical Physics.
  
  The unitary matrix $Q$ by Makhlin giving the isomorphism as an adjoint 
  action is studied and generalized from a different point of view. Some 
  problems are also presented.
  
  In particular, the homogeneous manifold $SU(2n)/SO(2n)$ which characterizes 
  entanglements in the case of $n=2$ is studied, and a clear--cut calculation  
  of the universal Yang--Mills action in (hep-th/0602204) is given for 
  the abelian case.
\end{abstract}
%


%
%
%
%

\section{Introduction}

The purpose of this paper is to reconsider the Makhlin's theorem and 
to generalize it in any qubit system, and moreover to apply to some 
subjects in Quantum Computation and Mathematical Physics. 

The isomorphism 
\[
SU(2)\otimes SU(2)\cong SO(4)
\]
is one of well--known theorems in elementary representation theory and is a 
typical characteristic of four dimensional euclidean space. However, it is 
usually abstract, see for example \cite{YS} \footnote{This is the most 
standard textbook in Japan on elementary representation theory}.

In \cite{YMa} Makhlin gave an interesting expression to the theorem. That is,
\[
F : SU(2)\otimes SU(2) \longrightarrow SO(4),\quad 
F(A\otimes B)=Q^{\dagger}(A\otimes B)Q
\]
with some unitary matrix $Q\in U(4)$. As far as we know this is the first that 
the map was given by {\bf the adjoint action}. The construction gave and will 
give many applications to both Quantum Computation and Mathematical Physics, 
see for example \cite{ZVWS} or \cite{RZ}.

In this paper we reconsider the construction (namely, $Q$) from a different 
point of view. Its construction is based on the Bell bases of 2--qubit system,  
so we treat a more general unitary matrix $R$ based on them. Our method may be 
clear and fresh.

Next we consider the problem whether or not it is possible to 
construct {\bf an inclusion}
\[
F : SU(2)\otimes SU(2)\otimes SU(2) \longrightarrow SO(8),\quad 
F(A\otimes B\otimes C)=R^{\dagger}(A\otimes B\otimes C)R
\]
with some unitary matrix $R\in U(8)$. A trial is made.

Since $SU(2)\otimes SU(2)\cong SO(4)$ entangled states in 2--qubit system  
are characterized by the homogeneous space $SU(4)/SO(4)$ called the 
Lagrangean Grassmannian. In Geometry it is generalized to $SU(n)/SO(n)$. 
We moreover enlarge it to $U(n)/O(n)$, which is isomorphic to the product 
space $U(1)\times SU(n)/SO(n)$. Here $U(1)$ is a kind of phase of $SU(n)/SO(n)$. 
We give an interesting coordinate system to $U(n)/O(n)$, which is not known 
as far as we know.

In last we apply to so--called universal Yang--Mills action being 
developed by us, \cite{FOS}. This is another non--linear generalization 
of usual Yang--Mills action \cite{YM}, which is different from the 
Born--Infeld one \cite{BI}. To write down the action form explicitly is 
not so easy due to its nonlinearity. We give a clear--cut derivation to it 
for the abelian case.

\section{The Isomorphism Revisited}

In this section we review the result in \cite{YMa} from a different point 
of view.

The 1--qubit space is $\fukuso^{2}=\mbox{Vect}_{\fukuso}\{\ket{0},\ket{1}\}$ 
where
\begin{equation}
\label{eq:bra-ket}
\ket{0}=
\left(
\begin{array}{c}
 1 \\
 0
\end{array}
\right),
\quad 
\ket{1}=
\left(
\begin{array}{c}
 0 \\
 1
\end{array}
\right).
\end{equation}
Let $\{\sigma_{1}, \sigma_{2}, \sigma_{3}\}$ be the Pauli matrices : 
\begin{equation}
\label{eq:Pauli matrices}
\sigma_{1} = 
\left(
  \begin{array}{cc}
    0 & 1 \\
    1 & 0
  \end{array}
\right), \quad 
\sigma_{2} = 
\left(
  \begin{array}{cc}
    0 & -i \\
    i & 0
  \end{array}
\right), \quad 
\sigma_{3} = 
\left(
  \begin{array}{cc}
    1 & 0 \\
    0 & -1
  \end{array}
\right).
\end{equation}
These ones act on the 1--qubit space.

Let us prepare some notations for the latter convenience. By $H(n;\fukuso)$ 
(resp. $H_{0}(n;\fukuso)$) the set of all (resp. all traceless) 
hermite matrices in $M(n;\fukuso)$
\[
H(n;\fukuso)=\{A\in M(n;\fukuso)\ |\ A^{\dagger}=A\}\ \supset\ 
H_{0}(n;\fukuso)=\{A\in H(n;\fukuso)\ |\ \mbox{tr}A=0\}.
\]
In particular, we have
\[
H_{0}(2;\fukuso)=\{a\equiv a_{1}\sigma_{1}+a_{2}\sigma_{2}+a_{3}\sigma_{3}\ 
|\ a_{1},a_{2},a_{3}\in \real\}.
\]
Here $H(n;\fukuso)\supset H(n,\real)$ is of course the set of all real  
symmetric matrices.

Next, let us consider the 2--qubit space 
$\fukuso^{2}\widehat{\otimes}\fukuso^{2}\cong \fukuso^{4}$, which is
\[
\fukuso^{2}\widehat{\otimes}\fukuso^{2}=\mbox{Vect}_{\fukuso}
\{\ket{00},\ket{01},\ket{10},\ket{11}\}
\]
where $\ket{ab}=\ket{a}\otimes \ket{b}\ (a,b\in \{0,1\})$. 

{\bf A comment is in order}.\ We in the following use notations on tensor 
product which are different from usual ones \footnote{We believe that our 
notations are clearer than usual ones}. That is, 
\[
\fukuso^{2}{\otimes}\fukuso^{2}=\{a\otimes b\ |\ a,b\in \fukuso^{2}\},
\]
while
\[
\fukuso^{2}\widehat{\otimes}\fukuso^{2}=
\left\{\sum_{j=1}^{k}c_{j}a_{j}\otimes b_{j}\ |\ a_{j},b_{j}\in \fukuso^{2},
\ c_{j}\in \fukuso,\ k\in \futon \right\}\cong \fukuso^{4}.
\]

\vspace{5mm}
We consider the Bell bases 
$\{\ket{\Psi_{1}},\ket{\Psi_{2}},\ket{\Psi_{3}},\ket{\Psi_{4}}\}$ 
which are defined as
\begin{eqnarray}
\label{eq:Bell bases}
\ket{\Psi_{1}}&=&\frac{1}{\sqrt{2}}(\ket{00}+\ket{11}),\quad
\ket{\Psi_{2}}=\frac{1}{\sqrt{2}}(\ket{01}+\ket{10}), \nonumber \\
\ket{\Psi_{3}}&=&\frac{1}{\sqrt{2}}(\ket{01}-\ket{10}),\quad
\ket{\Psi_{4}}=\frac{1}{\sqrt{2}}(\ket{00}-\ket{11}).
\end{eqnarray}
By making use of the bases the unitary matrix $R$ is defined as
\[
R=
\left(
\mbox{e}^{i\theta_{1}}\ket{\Psi_{1}},
\mbox{e}^{i\theta_{2}}\ket{\Psi_{2}},
\mbox{e}^{i\theta_{3}}\ket{\Psi_{3}},
\mbox{e}^{i\theta_{4}}\ket{\Psi_{4}}
\right)\ \in\ U(4).
\]
In the matrix form
\begin{equation}
\label{eq:matrix R}
R=\frac{1}{\sqrt{2}}
\left(
\begin{array}{cccc}
\mbox{e}^{i\theta_{1}} & 0 & 0 &  \mbox{e}^{i\theta_{4}} \\
0 & \mbox{e}^{i\theta_{2}} &  \mbox{e}^{i\theta_{3}} & 0 \\
0 & \mbox{e}^{i\theta_{2}} & -\mbox{e}^{i\theta_{3}} & 0 \\
\mbox{e}^{i\theta_{1}} & 0 & 0 & -\mbox{e}^{i\theta_{4}} 
\end{array}
\right).
\end{equation}

\vspace{3mm}
It is well--known the isomorphism
\[
F : SU(2)\otimes SU(2) \cong SO(4).
\]
To realize it as an adjoint action by $R$ (if it is possible)
\begin{equation}
\label{eq:Lie group level}
F(A\otimes B)=R^{\dagger}(A\otimes B)R \in SO(4),
\end{equation}
we have only to determine $\{\mbox{e}^{i\theta_{1}},\mbox{e}^{i\theta_{2}},
\mbox{e}^{i\theta_{3}},\mbox{e}^{i\theta_{4}}\}$ the coefficients of $R$. 
Let us consider this problem in a Lie algebra level because it is in general 
not easy to treat it in a Lie group level.

%
\begin{center}
\input{Lie-diagram.fig}
\end{center}

\vspace{5mm}
Since the Lie algebra of $SU(2)\otimes SU(2)$ is
\[
\mathfrak{L}(SU(2)\otimes SU(2))=
\left\{i(a\otimes 1_{2}+1_{2}\otimes b)\ |\ a,b \in H_{0}(2;\fukuso)\right\},
\]
we have only to examine
\begin{equation}
\label{eq:Lie algebra level}
f(i(a\otimes 1_{2}+1_{2}\otimes b))=
iR^{\dagger}(a\otimes 1_{2}+1_{2}\otimes b)R\in \mathfrak{L}(SO(4)).
\end{equation}
By setting $a=\sum_{j=1}^{3}a_{j}\sigma_{j}$ and 
$b=\sum_{j=1}^{3}b_{j}\sigma_{j}$ let us calculate the right hand side of 
(\ref{eq:Lie algebra level}). The result is 
\begin{eqnarray}
\label{eq:2-system relations}
&&iR^{\dagger}(a\otimes 1_{2}+1_{2}\otimes b)R= \nonumber \\
&&\left(
\begin{array}{cccc}
0 & 
i\mbox{e}^{-i(\theta_{1}-\theta_{2})}(a_{1}+b_{1}) & 
-\mbox{e}^{-i(\theta_{1}-\theta_{3})}(a_{2}-b_{2}) & 
i\mbox{e}^{-i(\theta_{1}-\theta_{4})}(a_{3}+b_{3})   \\
i\mbox{e}^{i(\theta_{1}-\theta_{2})}(a_{1}+b_{1}) & 
0 & 
i\mbox{e}^{-i(\theta_{2}-\theta_{3})}(a_{3}-b_{3}) & 
-\mbox{e}^{-i(\theta_{2}-\theta_{4})}(a_{2}+b_{2})   \\
 \mbox{e}^{i(\theta_{1}-\theta_{3})}(a_{2}-b_{2}) & 
i\mbox{e}^{i(\theta_{2}-\theta_{3})}(a_{3}-b_{3}) & 
0 & 
-i\mbox{e}^{-i(\theta_{3}-\theta_{4})}(a_{1}-b_{1})  \\
i\mbox{e}^{i(\theta_{1}-\theta_{4})}(a_{3}+b_{3})  & 
 \mbox{e}^{i(\theta_{2}-\theta_{4})}(a_{2}+b_{2})  & 
-i\mbox{e}^{i(\theta_{3}-\theta_{4})}(a_{1}-b_{1}) &
0 
\end{array}
\right).
\end{eqnarray}

Here if we set
\[
i\mbox{e}^{-i(\theta_{1}-\theta_{2})}=1,\ 
i\mbox{e}^{-i(\theta_{1}-\theta_{4})}=1,\
i\mbox{e}^{-i(\theta_{2}-\theta_{3})}=1,\
-i\mbox{e}^{-i(\theta_{3}-\theta_{4})}=1,
\]
from which $-\mbox{e}^{-i(\theta_{1}-\theta_{3})}=1$ and 
$-\mbox{e}^{-i(\theta_{2}-\theta_{4})}=-1$ automatically, then we have 
\[
\mbox{e}^{i\theta_{1}}=1,\ \mbox{e}^{i\theta_{2}}=-i,\
\mbox{e}^{i\theta_{3}}=-1,\ \mbox{e}^{i\theta_{4}}=-i.
\]
Therefore our $R$ becomes
\begin{equation}
\label{eq:R not Q}
R=\frac{1}{\sqrt{2}}
\left(
  \begin{array}{cccc}
    1 &  0 &  0 & -i  \\
    0 & -i & -1 &  0  \\
    0 & -i &  1 &  0  \\
    1 &  0 &  0 &  i 
  \end{array}
\right).
\end{equation}
We used the notation $R$ again for simplicity. 
Note that the unitary matrix $R$ is a bit different from $Q$ in \cite{YMa}.

For the latter convenience let us rewrite. If we set
\begin{eqnarray}
\label{eq:correspondence}
iR^{\dagger}(a\otimes 1_{2}+1_{2}\otimes b)R 
&=&
\left(
  \begin{array}{cccc}
    0 & a_{1}+b_{1} & a_{2}-b_{2} & a_{3}+b_{3}       \\
    -(a_{1}+b_{1}) & 0 & a_{3}-b_{3} & -(a_{2}+b_{2}) \\
    -(a_{2}-b_{2}) & -(a_{3}-b_{3}) & 0 & a_{1}-b_{1} \\
    -(a_{3}+b_{3}) & a_{2}+b_{2} & -(a_{1}-b_{1}) & 0  
  \end{array}
\right)  \nonumber \\
&\equiv&
\left(
  \begin{array}{cccc}
    0 & f_{12} & f_{13} & f_{14}   \\
   -f_{12} & 0 & f_{23} & f_{24}   \\
   -f_{13} & -f_{23} & 0 & f_{34}  \\
   -f_{14} & -f_{24} & -f_{34} & 0  
  \end{array}
\right)\in \mathfrak{L}(SO(4))
\end{eqnarray}
then we obtain
\begin{eqnarray}
\label{eq:left-a}
&&a=a_{1}\sigma_{1}+a_{2}\sigma_{2}+a_{3}\sigma_{3}
   =\frac{f_{12}+f_{34}}{2}\sigma_{1}+
    \frac{f_{13}-f_{24}}{2}\sigma_{2}+
    \frac{f_{14}+f_{23}}{2}\sigma_{3}, \\
\label{eq:right-b}
&&b=b_{1}\sigma_{1}+b_{2}\sigma_{2}+b_{3}\sigma_{3}
   =\frac{f_{12}-f_{34}}{2}\sigma_{1}-
    \frac{f_{13}+f_{24}}{2}\sigma_{2}+
    \frac{f_{14}-f_{23}}{2}\sigma_{3}.
\end{eqnarray}

\section{A Trial toward Generalization}

We would like to generalize the result $SU(2)\otimes SU(2)\cong SO(4)$ 
in the preceding section. Of course it is not true that 
$SU(2)\otimes SU(2)\otimes SU(2)\cong SO(8)$. Our question is as follows : 
is it possible to find {\bf an inclusion}
\[
F : SU(2)\otimes SU(2)\otimes SU(2)\ \longrightarrow\ SO(8)
\]
with the form 
\begin{equation}
F(A\otimes B\otimes C)=R^{\dagger}(A\otimes B\otimes C)R\in SO(8)
\end{equation}
by finding a unitary matrix $R\in U(8)$ ?

Let us make a trial in the following. In the preceding section we used the 
Bell bases to construct the unitary matrix $R$, so in this case we trace 
the same line \footnote{We believe the way of thinking natural}. 

In 3--qudit system the generalized ``Bell bases" are known to be \\
$\{\ket{\Psi_{1}},\ket{\Psi_{2}},\ket{\Psi_{3}},\ket{\Psi_{4}},
\ket{\Psi_{5}},\ket{\Psi_{6}},\ket{\Psi_{7}},\ket{\Psi_{8}}\}$ where
\begin{eqnarray}
\label{eq:three Bell bases}
\ket{\Psi_{1}}&=&\frac{1}{\sqrt{2}}(\ket{000}+\ket{111}),\quad
\ket{\Psi_{2}}=\frac{1}{\sqrt{2}}(\ket{001}+\ket{110}), \nonumber \\
\ket{\Psi_{3}}&=&\frac{1}{\sqrt{2}}(\ket{010}+\ket{101}),\quad
\ket{\Psi_{4}}=\frac{1}{\sqrt{2}}(\ket{011}+\ket{100}), \nonumber \\
\ket{\Psi_{5}}&=&\frac{1}{\sqrt{2}}(\ket{011}-\ket{100}),\quad 
\ket{\Psi_{6}}=\frac{1}{\sqrt{2}}(\ket{010}-\ket{101}), \nonumber \\
\ket{\Psi_{7}}&=&\frac{1}{\sqrt{2}}(\ket{001}-\ket{110}),\quad
\ket{\Psi_{8}}=\frac{1}{\sqrt{2}}(\ket{000}-\ket{111}),
\end{eqnarray}
see for example \cite{GT}. Then the unitary matrix $R$ corresponding to 
(\ref{eq:matrix R}) is given by
\[
R=
\left(
\mbox{e}^{i\theta_{1}}\ket{\Psi_{1}},
\mbox{e}^{i\theta_{2}}\ket{\Psi_{2}},
\mbox{e}^{i\theta_{3}}\ket{\Psi_{3}},
\mbox{e}^{i\theta_{4}}\ket{\Psi_{4}},
\mbox{e}^{i\theta_{5}}\ket{\Psi_{5}},
\mbox{e}^{i\theta_{6}}\ket{\Psi_{6}},
\mbox{e}^{i\theta_{7}}\ket{\Psi_{7}},
\mbox{e}^{i\theta_{8}}\ket{\Psi_{8}}
\right)\ \in\ U(8),
\]
or in the matrix form 
\begin{equation}
\label{eq:matrix R in three}
R=\frac{1}{\sqrt{2}}
\left(
\begin{array}{cccccccc}
\mbox{e}^{i\theta_{1}} & 0 & 0 & 0 & 0 & 0 & 0 &  \mbox{e}^{i\theta_{8}} \\
0 & \mbox{e}^{i\theta_{2}} & 0 & 0 & 0 & 0 &  \mbox{e}^{i\theta_{7}} & 0 \\
0 & 0 & \mbox{e}^{i\theta_{3}} & 0 & 0 &  \mbox{e}^{i\theta_{6}} & 0 & 0 \\
0 & 0 & 0 & \mbox{e}^{i\theta_{4}} &  \mbox{e}^{i\theta_{5}} & 0 & 0 & 0 \\
0 & 0 & 0 & \mbox{e}^{i\theta_{4}} & -\mbox{e}^{i\theta_{5}} & 0 & 0 & 0 \\
0 & 0 & \mbox{e}^{i\theta_{3}} & 0 & 0 & -\mbox{e}^{i\theta_{6}} & 0 & 0 \\
0 & \mbox{e}^{i\theta_{2}} & 0 & 0 & 0 & 0 & -\mbox{e}^{i\theta_{7}} & 0 \\
\mbox{e}^{i\theta_{1}} & 0 & 0 & 0 & 0 & 0 & 0 & -\mbox{e}^{i\theta_{8}}
\end{array}
\right).
\end{equation}

We must check whether or not it is possible to construct
\begin{equation}
F(A\otimes B\otimes C)=R^{\dagger}(A\otimes B\otimes C)R \in SO(8)
\end{equation}
by determining the coefficients $\{\mbox{e}^{i\theta_{1}},
\mbox{e}^{i\theta_{2}},\mbox{e}^{i\theta_{3}},\mbox{e}^{i\theta_{4}},
\mbox{e}^{i\theta_{5}},\mbox{e}^{i\theta_{6}},\mbox{e}^{i\theta_{7}},
\mbox{e}^{i\theta_{8}}\}$.

Similarly in the preceding section we have only to check it 
in a Lie algebra level :
\[
f(i(a\otimes 1_{2}\otimes 1_{2}+1_{2}\otimes b\otimes 1_{2}+
1_{2}\otimes 1_{2}\otimes c))=
iR^{\dagger}(a\otimes 1_{2}\otimes 1_{2}+1_{2}\otimes b\otimes 1_{2}+
1_{2}\otimes 1_{2}\otimes c)R\in \mathfrak{L}(SO(8)).
\]

The result is negative. That is, the coefficients $\{\mbox{e}^{i\theta_{1}},
\mbox{e}^{i\theta_{2}},\mbox{e}^{i\theta_{3}},\mbox{e}^{i\theta_{4}},
\mbox{e}^{i\theta_{5}},\mbox{e}^{i\theta_{6}},\mbox{e}^{i\theta_{7}},
\mbox{e}^{i\theta_{8}}\}$ satisfying the above equation 
don't exist (a long calculation like (\ref{eq:2-system relations}) is omitted). 
Therefore we again propose

\par \vspace{5mm} \noindent
{\bf Problem}\quad Does a unitary matrix $R\in SU(8)$ exist giving 
an inclusion 
\[
F : SU(2)\otimes SU(2)\otimes SU(2)\ \longrightarrow\ SO(8),\quad 
F(A\otimes B\otimes C)=R^{\dagger}(A\otimes B\otimes C)R\ \ ?
\]

\vspace{5mm}
A comment is in order. If we can find such an inclusion then we have 
the following fiber bundle
\[
\frac{SO(8)}{F(SU(2)\otimes SU(2)\otimes SU(2))}\ 
\longrightarrow\
\frac{SU(8)}{F(SU(2)\otimes SU(2)\otimes SU(2))}\ 
\longrightarrow\ 
\frac{SU(8)}{SO(8)}.
\]
That is, entangled states for 3--qubit system are characterized by the 
homogeneous space $SU(8)/F(SU(2)\otimes SU(2)\otimes SU(2))$ and 
this space is understood by the fiber bundle.

\section{General R Matrix}

In this section we treat the general $n$--qubit system. 
We would like to generalize the unitary matrix $R$ in the preceding sections.
Generalized ``Bell bases" are constructed as follows. 

\par \noindent
For $0\leq k\leq 2^{n-1}-1$, since $k$ can be written as
\[
k=a_{0}2^{n-2}+a_{1}2^{n-3}+\cdots+a_{n-3}2+a_{n-2},\quad 
a_{j}\in \{0,1\}
\]
we set
\begin{eqnarray}
\ket{\Psi_{k+1}}&=&\frac{1}{\sqrt{2}}
\left\{\ket{0a_{0}a_{1}\cdots a_{n-2}}+
\ket{1{\breve a}_{0}{\breve a}_{1}\cdots {\breve a}_{n-2}}\right\}, 
\nonumber \\
\ket{\Psi_{2^{n}-k}}&=&\frac{1}{\sqrt{2}}
\left\{\ket{0a_{0}a_{1}\cdots a_{n-2}}-
\ket{1{\breve a}_{0}{\breve a}_{1}\cdots {\breve a}_{n-2}}\right\}
\end{eqnarray}
where ${\breve a}_{j}=1-a_{j}$. 

We define the unitary matrix $R$ as
\[
R=
\left(
\mbox{e}^{i\theta_{1}}\ket{\Psi_{1}},
\mbox{e}^{i\theta_{2}}\ket{\Psi_{2}},
\cdots,
\mbox{e}^{i\theta_{2^{n-1}}}\ket{\Psi_{2^{n-1}}},
\mbox{e}^{i\theta_{2^{n-1}+1}}\ket{\Psi_{2^{n-1}+1}},
\cdots,
\mbox{e}^{i\theta_{2^{n}-1}}\ket{\Psi_{2^{n}-1}},
\mbox{e}^{i\theta_{2^{n}}}\ket{\Psi_{2^{n}}}
\right).
\]
In the matrix form
\begin{equation}
\label{eq:matrix R in general}
R=\frac{1}{\sqrt{2}}
\left(
\begin{array}{cccccccc}
\mbox{e}^{i\theta_{1}} &&&&&&& \mbox{e}^{i\theta_{2^{n}}} \\
& \mbox{e}^{i\theta_{2}} &&&&& \mbox{e}^{i\theta_{2^{n}-1}} & \\
&& \ddots &&& \rddots && \\
&&& \mbox{e}^{i\theta_{2^{n-1}}} & \mbox{e}^{i\theta_{2^{n-1}+1}} &&& \\
&&& \mbox{e}^{i\theta_{2^{n-1}}} & -\mbox{e}^{i\theta_{2^{n-1}+1}} &&& \\
&& \rddots &&& \ddots && \\
& \mbox{e}^{i\theta_{2}} &&&&& -\mbox{e}^{i\theta_{2^{n}-1}} & \\
\mbox{e}^{i\theta_{1}} &&&&&&& -\mbox{e}^{i\theta_{2^{n}}}
\end{array}
\right).
\end{equation}
For $n=2$ and $3$ the matrix $R$ in the preceding sections is recovered.

Now we give a characterization to $R$. That is, $R$ satisfies the equation
\begin{equation}
R(\sigma_{3}\otimes 1_{2}\otimes \cdots \otimes 1_{2})R^{\dagger}=
\sigma_{1}\otimes \sigma_{1}\otimes \cdots \otimes \sigma_{1}.
\end{equation}
The proof is left to readers (check this for (\ref{eq:matrix R}) and 
(\ref{eq:matrix R in three})).

If we define 
\[
\widetilde{W}=W\otimes 1_{2}\otimes \cdots \otimes 1_{2}
\]
where $W$ is the usual Walsh--Hadamard matrix ($W\sigma_{3}W=\sigma_{1}$), 
then the product $R\widetilde{W}$ gives
\begin{equation}
R\widetilde{W}(\sigma_{1}\otimes 1_{2}\otimes \cdots \otimes 1_{2})
(R\widetilde{W})^{\dagger}=
\sigma_{1}\otimes \sigma_{1}\otimes \cdots \otimes \sigma_{1}.
\end{equation}
That is, $R\widetilde{W}$ is a copy operation for $\sigma_{1}$ :
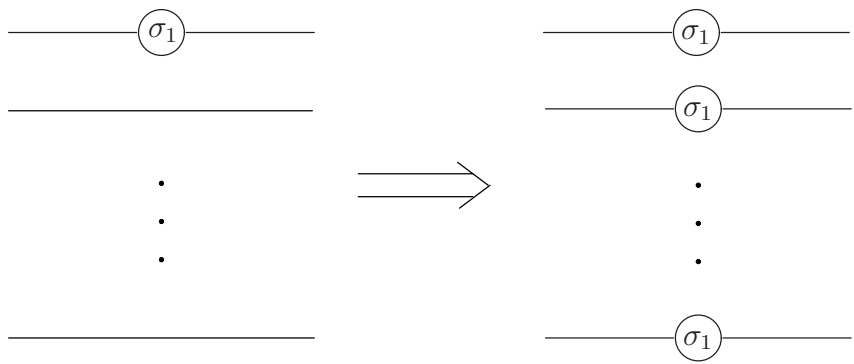
\vspace{5mm}
\begin{center}
\input{copy-sigma.fig}
\end{center}

We believe that $R$ will play an important role in Quantum Computation.

\section{Application to Elementary Represetation Theory}

To construct a representation from $SU(2)$ to $SO(3)$ is very well--known. 
In this section we point out a relation between the representation $\rho$ 
and the isomorphism $F$ in the section 2.

First of all let us make a review of constructing the representation,
\begin{equation}
\rho : SU(2)\ {\longrightarrow}\ SO(3),
\end{equation}
see for example \cite{KF3}, Appendix F.\ For the matrix
\[
g=
\left(
 \begin{array}{cc}
  a+ib & c+id \\
 -c+id & a-ib
  \end{array}
\right)
\quad a, b, c, d\ \in \real
\]
it is easy to see
\[
g \in SU(2) \Longleftrightarrow a^{2}+b^{2}+c^{2}+d^{2}=1. 
\]
For the Pauli matrices $\{\sigma_{1},\sigma_{2},\sigma_{3}\}$ in 
(\ref{eq:Pauli matrices}) we set 
\[
\tau_{j}=\frac{1}{2}\sigma_{j} \quad \mbox{for}\quad j=1,2,3.
\]
Then the representation $\rho$ is constructed as follows : since
\begin{eqnarray}
g\tau_{1}g^{-1}&=&(a^{2}-b^{2}-c^{2}+d^{2})\tau_{1}
-2(ab-cd)\tau_{2}+2(ac+bd)\tau_{3}, \nonumber \\
g\tau_{2}g^{-1}&=&2(ab+cd)\tau_{1}
+(a^{2}-b^{2}+c^{2}-d^{2})\tau_{2}-2(ad-bc)\tau_{3}, \nonumber \\
g\tau_{3}g^{-1}&=&-2(ac-bd)\tau_{1}
+2(ad+bc)\tau_{2}+(a^{2}+b^{2}-c^{2}-d^{2})\tau_{3}, \nonumber 
\end{eqnarray}
we have 
\[
\left(g\tau_{1}g^{-1}, g\tau_{2}g^{-1}, g\tau_{3}g^{-1}\right)
=\left(\tau_{1}, \tau_{2}, \tau_{3}\right)\rho(g)
\]
where 
\begin{equation}
\rho(g)=
\left(
 \begin{array}{ccc}
  a^{2}-b^{2}-c^{2}+d^{2}& 2(ab+cd)& -2(ac-bd) \\
  -2(ab-cd)& a^{2}-b^{2}+c^{2}-d^{2}& 2(ad+bc) \\
  2(ac+bd)& -2(ad-bc)& a^{2}+b^{2}-c^{2}-d^{2}
  \end{array}
\right).
\end{equation}
Here let us rewrite the above. Noting that 
\begin{eqnarray}
a^{2}-b^{2}-c^{2}+d^{2}&=&1-2(b^{2}+c^{2}),\quad
a^{2}-b^{2}+c^{2}-d^{2}=1-2(b^{2}+d^{2}),  \nonumber \\
a^{2}+b^{2}-c^{2}-d^{2}&=&1-2(c^{2}+d^{2}),  \nonumber 
\end{eqnarray}
from $a^{2}+b^{2}+c^{2}+d^{2}=1$, we obtain
\begin{equation}
\rho(g)=
\left(
 \begin{array}{ccc}
  1-2(b^{2}+c^{2})& 2(ab+cd)& -2(ac-bd) \\
  -2(ab-cd)& 1-2(b^{2}+d^{2})& 2(ad+bc) \\
  2(ac+bd)& -2(ad-bc)& 1-2(c^{2}+d^{2})
 \end{array}
\right).
\end{equation}

Now we would like to search a relation induced from $\rho$ and $F$. 
See the following diagram : 
\vspace{3mm}
\begin{center}
 \input{inclusion.fig}
\end{center}
\vspace{3mm}
where $\iota$ is the natural inclusion defined by 
\[
\iota(O)=
\left(
 \begin{array}{cc}
  1 &   \\
    & O
 \end{array}
\right)\in SO(4)\quad \mbox{for}\quad O\in SO(3).
\]
We determine the map $\tilde{\iota}$ defined by 
\[
\tilde{\iota}=F^{-1}\circ \iota\circ \rho \ \Longrightarrow\ 
\tilde{\iota}(g)=R
\left(
 \begin{array}{cc}
  1 &         \\
    & \rho(g)
 \end{array}
\right)R^{\dagger}\quad \mbox{for}\quad g\in SU(2).
\]

The result is
\begin{equation}
\label{eq:natural-map?}
\tilde{\iota}\left(
\left(
 \begin{array}{cc}
  a+ib&  c+id \\
  -c+id& a-ib
  \end{array}
\right)
\right)
=
\left(
 \begin{array}{cc}
  a+ib & c+id \\
 -c+id & a-ib
  \end{array}
\right)
\otimes
\left(
 \begin{array}{cc}
  a-ib & c-id \\
 -c-id & a+ib
  \end{array}
\right).
\end{equation}
Let us rewrite the equation. It is easy to see
\[
\mbox{The right hand side of (\ref{eq:natural-map?})}
=g\otimes \sigma_{2}g\sigma_{2}^{\dagger}
=(1_{2}\otimes \sigma_{2})(g\otimes g)(1_{2}\otimes \sigma_{2})^{\dagger},
\]
so we have
\begin{equation}
\label{eq:iota-map}
\tilde{\iota}(g)=
(1_{2}\otimes \sigma_{2})(g\otimes g)(1_{2}\otimes \sigma_{2})^{\dagger}
\equiv 
(1_{2}\otimes \sigma_{2})\Delta(g)(1_{2}\otimes \sigma_{2})^{\dagger}.
\end{equation}

\vspace{3mm}
{\bf A comment is in order}.\ If we rewrite (\ref{eq:iota-map}) as
\[
\left(
 \begin{array}{cc}
  1 &         \\
    & \rho(g)
 \end{array}
\right)
\equiv 1\oplus \rho(g)=
R^{\dagger}(1_{2}\otimes \sigma_{2})(g\otimes g)
\left\{R^{\dagger}(1_{2}\otimes \sigma_{2})\right\}^{\dagger},
\]
this is the well--known irreducible decomposition of tensor product\ 
$\frac{1}{2}\otimes \frac{1}{2}=0\oplus 1$\ 
in terms of spin representation, compare this with \cite{FHKSW}.

\section{Lagrangean Grassmannians}

Since $SU(2)\otimes SU(2)\cong SO(4)$, entangled states in 2--qubit system 
are characterized by the homogeneous space $SU(4)/SO(4)$, which is a special 
case of the Lagrangean Grassmannians $SU(n)/SO(n)\ \mbox{for}\ n\geq 2$. 
Here let us expand $SU(n)/SO(n)$ a little bit, namely our target is 
$U(n)/O(n)$. See for example \cite{KF2} for the Grassmannians.

For $n\geq 1$ we define the two sets (spaces)
\begin{eqnarray}
X_{n}&\equiv& \left\{R\in U(n)\ |\ R^{T}=R \right\}, \\
\widetilde{X}_{n}&\equiv& \left\{R\in SU(n)\ |\ R^{T}=R \right\}.
\end{eqnarray}
Then it is easy to see
\begin{eqnarray}
X_{n}&=&\left\{A^{T}A \ |\ A\in U(n) \right\}\cong \frac{U(n)}{O(n)}, \\
\widetilde{X}_{n}&=&\left\{A^{T}A \ |\ A\in SU(n) \right\}\cong 
\frac{SU(n)}{SO(n)}
\end{eqnarray}
and
\begin{equation}
X_{n}\cong U(1)\times \widetilde{X}_{n}
\end{equation}
because $R\in X_{n}$ can be written as $R=\mbox{e}^{i\theta}\tilde{R}$ 
where $\tilde{R}\in \widetilde{X}_{n}$.

We can give a very interesting coordinate system (as a manifold) to $X_{n}$
\footnote{However, it is not easy for $\widetilde{X}_{n}$}.

\par \noindent
Let $X_{n}\ni R_{0}=A^{T}A$ be fixed and define a neighborhood of $R_{0}$ as
\[
U_{A}=\left\{R\in X_{n}\ |\ R+R_{0}\in GL(n;\fukuso)\right\}
\]
and define a map
\begin{equation}
\phi_{A} : U_{A}\ \longrightarrow\ H(n;\real),\quad
\phi_{A}(R)=2i\left(E+\bar{A}R\bar{A}^{T}\right)^{-1}
\left(E-\bar{A}R\bar{A}^{T}\right).
\end{equation}
It is easy to see
\[
\phi_{A}(R)=2i\left\{2\left(E+\bar{A}R\bar{A}^{T}\right)^{-1}-E\right\}.
\]

\par \noindent
Conversely, we define a map
\begin{equation}
\omega_{A} : H(n;\real)\ \longrightarrow\ U_{A},\quad
\omega_{A}(X)=A^{T}\left(E+\frac{i}{2}X\right)
\left(E-\frac{i}{2}X\right)^{-1}A.
\end{equation}
Then it is not difficult to see
\[
\omega_{A}\circ \phi_{A}=1_{U_{A}}\quad \mbox{and}\quad
\phi_{A}\circ \omega_{A}=1_{H(n;\real)}.
\]

Next, for $U_{A}\cap U_{B}\ni R$ satisfying $R=\omega_{A}(X)=\omega_{B}(Y)$ 
it is not difficult to solve
\begin{equation}
Y=2\left(\alpha-\frac{1}{2}X\beta\right)^{-1}
\left(\beta+\frac{1}{2}X\alpha\right)
\ \Longleftrightarrow\ 
Y=\phi_{B}\circ \phi_{A}^{-1}(X)
\end{equation}
if we set $AB^{\dagger}=\alpha+i\beta$. See the following diagram :

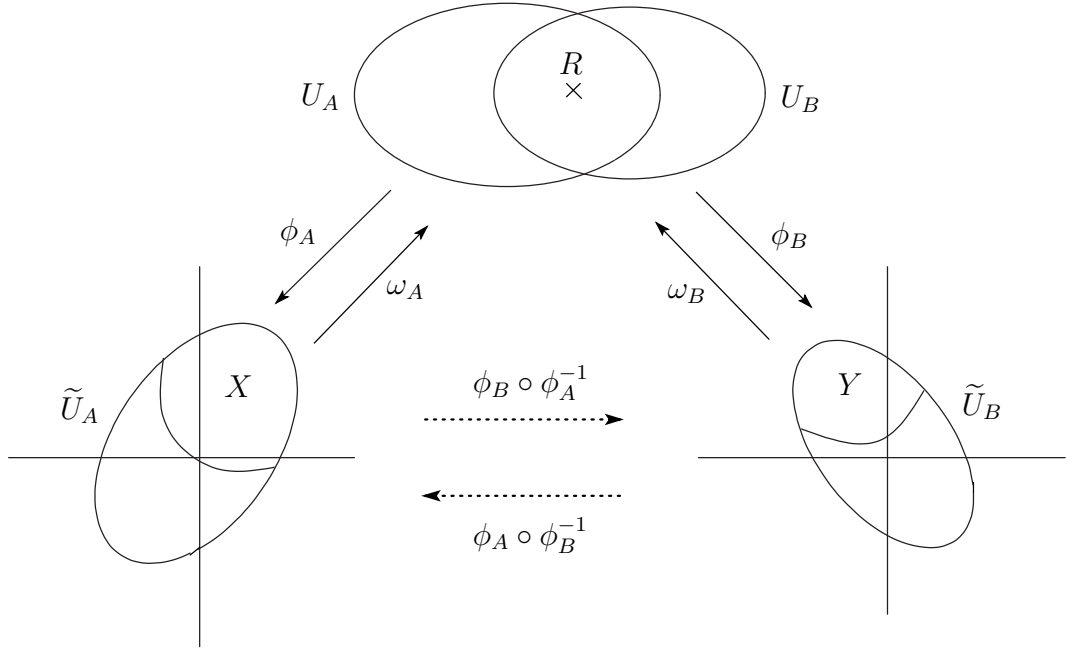
\vspace{5mm}
\begin{center}
\input{manifold.fig}
\end{center}

The aim of the paper is not to study detailed (geometric) structures of 
$X_{n}\cong U(n)/O(n)$, for the case of $n=4$ especially,  
so we leave it in a forthcoming paper.

\section{Application to Universal Yang--Mills Action}

In this section we revisit \cite{FOS} from a different point of view. 
In the following we consider the ${\bf C}^{\infty}$ category, namely 
${\bf C}^{\infty}$--manifolds, ${\bf C}^{\infty}$--maps, etc. 
See \cite{MN} for more detailed descriptions. 

Let $M$ be a four dimensional manifold and $G$ a classical group, 
in particular, $U(1)$ and $SU(2)$. We denote by $\{G, P, \pi, M\}$ 
a principal $G$ bundle on $M$
\[
\pi : P\ \longrightarrow\ M,\quad \pi^{-1}(m)\cong G.
\]

We consider a theory of connections of the principal $G$ bundle, so let 
$g\equiv \mathfrak{L}(G)$ be 
the Lie algebra of the group $G$. In the following we treat it locally, 
which is enough for our purpose. 
Let $U$ be any open set in $M$, then a connection $\{A_{\mu}\}$ (a gauge 
potential) is
\[
A_{\mu} : U\ \longrightarrow\ g
\]
and the corresponding curvature $\{F_{\mu\nu}\}$ is given by
\[
F_{\mu\nu}
= \frac{\partial A_{\nu}}{\partial x_{\mu}}-
\frac{\partial A_{\mu}}{\partial x_{\nu}}+[A_{\mu},A_{\nu}]
\equiv \partial_{\mu}A_{\nu}-\partial_{\nu}A_{\mu}+[A_{\mu},A_{\nu}].
\]
We note that 
\[
F_{\mu\nu} : U\ \longrightarrow\ g,\qquad F_{\nu\mu}=-F_{\mu\nu}.
\]

For a map $\phi : U\ \longrightarrow\ G$,\ a gauge transformation by $\phi$ 
is defined by
\[
A_{\mu}\ \longrightarrow\ \phi^{-1}A_{\mu}\phi +\phi^{-1}\partial_{\mu}\phi
\]
and the curvature is then transformed like 
\[
F_{\mu\nu}\ \longrightarrow\ \phi^{-1}F_{\mu\nu}\phi.
\]

In the following we consider only the abelian case $G=U(1)$. For that let us 
define a curvature matrix
\begin{equation}
\label{eq:abelian-curvature-matrix}
{\cal F}=
\left(
\begin{array}{cccc}
  0      & F_{12}  & F_{13}  & F_{14} \\
 -F_{12} & 0       & F_{23}  & F_{24} \\
 -F_{13} & -F_{23} & 0       & F_{34} \\
 -F_{14} & -F_{24} & -F_{34} & 0
\end{array}
\right).
\end{equation}

The abelian Yang--Mills action ${\cal A}_{YM}$ \footnote{
it may be suitable to call it the Maxwell action} 
is given by
\begin{equation}
\label{eq:aYM-action}
{\cal A}_{YM}\equiv \frac{1}{2}\mbox{tr} \left(g{\cal F}\right)^{2}
=-g^{2}\sum_{i<j}F_{ij}^{2}
\end{equation}
where $g$ is a coupling constant, see \cite{YM}.

This model is very well--known and there is nothing added furthermore. 
We are interested in some non--linear generalization(s) of it.

A non--linear extension of the abelian Yang--Mills is known as the 
Born--Infeld theory whose action ${\cal A}_{BI}$ is given by
\begin{equation}
\label{eq:BI-action}
 {\cal A}_{BI}\equiv \sqrt{\mbox{det}\left({\bf 1}_{4}+g{\cal F}\right)}
\end{equation}
where $g$ is a coupling constant, see \cite{BI}.

As to this model and its generalizations to non--abelian groups 
there are a lot of papers. However, we don't make a comment in the paper. 

On the other hand, we have presented another non--linear extension, 
\cite{FOS}. Its action ${\cal A}$ is given by
\begin{equation}
\label{eq:FOS-action}
{\cal A}_{FOS}\equiv \mbox{tr}\ \mbox{e}^{g{\cal F}}
\end{equation}
where $g$ is a coupling constant. We have called this the universal 
Yang--Mills action.

Let us calculate the right hand side of (\ref{eq:FOS-action}) by making use 
of the result in section 2, which is more clear--cut. 
From (\ref{eq:correspondence}) with (\ref{eq:left-a}) and (\ref{eq:right-b})
\[
\mbox{e}^{g{\cal F}}=R^{\dagger}R\mbox{e}^{g{\cal F}}R^{\dagger}R=
R^{\dagger}\mbox{e}^{gR{\cal F}R^{\dagger}}R,
\]
while
\[
R{\cal F}R^{\dagger}=i(a\otimes 1_{2}+1_{2}\otimes b)
\]
with
\begin{eqnarray}
&&a=\frac{F_{12}+F_{34}}{2}\sigma_{1}+
    \frac{F_{13}-F_{24}}{2}\sigma_{2}+
    \frac{F_{14}+F_{23}}{2}\sigma_{3}, \\
&&b=\frac{F_{12}-F_{34}}{2}\sigma_{1}-
    \frac{F_{13}+F_{24}}{2}\sigma_{2}+
    \frac{F_{14}-F_{23}}{2}\sigma_{3}.
\end{eqnarray}
Therefore
\[
\mbox{e}^{gR{\cal F}R^{\dagger}}=
\mbox{e}^{ig(a\otimes 1_{2}+1_{2}\otimes b)}=
\mbox{e}^{iga}\otimes \mbox{e}^{igb}
\]
and
\[
\mbox{tr}\ \mbox{e}^{g{\cal F}}=
\mbox{tr}\ \mbox{e}^{gR{\cal F}R^{\dagger}}=
\mbox{tr}\ \mbox{e}^{iga}\otimes \mbox{e}^{igb}=
\mbox{tr}\ \mbox{e}^{iga}\ \mbox{tr}\ \mbox{e}^{igb}.
\]
Noting the well--known formula
\[
\mbox{e}^{i(x\sigma_{1}+y\sigma_{2}+z\sigma_{3})}
=\cos{r} 1_{2}+\frac{\sin{r}}{r}i(x\sigma_{1}+y\sigma_{2}+z\sigma_{3}),\quad 
r\equiv \sqrt{x^{2}+y^{2}+z^{2}}
\]
we finally obtain
\begin{equation}
\label{eq:abelian-result}
{\cal A}_{FOS}\equiv \mbox{tr}\ \mbox{e}^{g{\cal F}}
=4\cos\left(gX_{asd}\right)\cos\left(gX_{sd}\right)
\end{equation}
where
\begin{eqnarray}
X_{sd}^{2}&=&\frac{1}{4}
\left\{(F_{12}-F_{34})^{2}+(F_{13}+F_{24})^{2}+(F_{14}-F_{23})^{2}\right\},\\
X_{asd}^{2}&=&\frac{1}{4}
\left\{(F_{12}+F_{34})^{2}+(F_{13}-F_{24})^{2}+(F_{14}+F_{23})^{2}\right\}.
\end{eqnarray}
It is very notable that
\begin{eqnarray*}
&&X_{sd}=0 \Longleftrightarrow 
F_{12}=F_{34},\ F_{13}=-F_{24},\ F_{14}=F_{23} \Longleftrightarrow 
\{F_{ij}\}\ \mbox{is {\bf self--dual}}, \\
&&X_{asd}=0 \Longleftrightarrow 
F_{12}=-F_{34},\ F_{13}=F_{24},\ F_{14}=-F_{23} \Longleftrightarrow 
\{F_{ij}\}\ \mbox{is {\bf anti--self--dual}}.
\end{eqnarray*}

The characteristic of our model is that the action (\ref{eq:abelian-result}) 
splits {\bf automatically} into two parts consisting of self--dual and 
anti--self--dual directions. 
Namely, we have {\bf automatically} the self--dual and anti--self--dual 
equations without solving the equations of motion as in a usual case. 

Last in this section let us make an important comment on the action 
(\ref{eq:abelian-result}). Since the target (as a map) of curvatures 
$\{F_{ij}\}$ is the abelian Lie algebra $u(1)=\sqrt{-1}{\bf R}$ it may be 
appropriate to write $F_{ij}=\sqrt{-1}G_{ij}$. 
Then (\ref{eq:abelian-result}) is changed into the more suitable form
\begin{equation}
\label{eq:abelian-result-modify}
{\cal A}_{FOS}=
4\cosh\left(gY_{asd}\right)\cosh\left(gY_{sd}\right)
\end{equation}
with
\begin{eqnarray}
Y_{sd}^{2}&=&\frac{1}{4}
\left\{(G_{12}-G_{34})^{2}+(G_{13}+G_{24})^{2}+(G_{14}-G_{23})^{2}\right\}
\geq\ 0, \\
Y_{asd}^{2}&=&\frac{1}{4}
\left\{(G_{12}+G_{34})^{2}+(G_{13}-G_{24})^{2}+(G_{14}+G_{23})^{2}\right\}
\geq\ 0.
\end{eqnarray}

\section{Discussion}

In this letter we studied several aspects coming from the isomorphism 
$SU(2)\otimes SU(2)\cong SO(4)$ in Quantum Computation and 
Mathematical Physics and moreover presented the problem concerning 
a possibility of generalization.  
Details of the paper and further developments will be published elsewhere. 

\vspace{3mm}
In last let us make a brief comment. We are studying a quantum computation 
based on Cavity QED and have presented a model, see \cite{FHKW1} and 
\cite{FHKW2} in detail. The image is as follows (the details are omitted) :

\vspace{5mm}
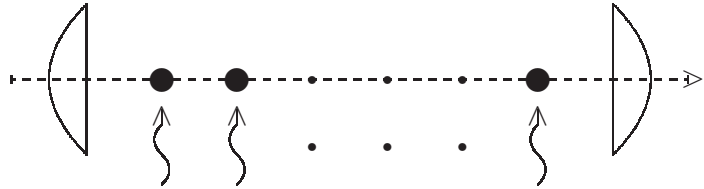
\begin{center}
\setlength{\unitlength}{1mm} 
\begin{picture}(110,40)(0,-20)
\bezier{200}(20,0)(10,10)(20,20)
\put(20,0){\line(0,1){20}}
\put(30,10){\circle*{3}}
\bezier{200}(30,-4)(32,-2)(30,0)
\bezier{200}(30,0)(28,2)(30,4)
\put(30,4){\line(0,1){2}}
\put(28.6,4){$\wedge$}
\put(40,10){\circle*{3}}
\bezier{200}(40,-4)(42,-2)(40,0)
\bezier{200}(40,0)(38,2)(40,4)
\put(40,4){\line(0,1){2}}
\put(38.6,4){$\wedge$}
\put(50,10){\circle*{1}}
\put(60,10){\circle*{1}}
\put(70,10){\circle*{1}}
\put(50,1){\circle*{1}}
\put(60,1){\circle*{1}}
\put(70,1){\circle*{1}}
\put(80,10){\circle*{3}}
\bezier{200}(80,-4)(82,-2)(80,0)
\bezier{200}(80,0)(78,2)(80,4)
\put(80,4){\line(0,1){2}}
\put(78.6,4){$\wedge$}
\bezier{200}(90,0)(100,10)(90,20)
\put(90,0){\line(0,1){20}}
\put(10,10){\dashbox(90,0)}
\put(99,9){$>$}
\end{picture}
\end{center}
\vspace{-20mm}
\begin{center}
{The general setting for a quantum computation based on Cavity QED : \\
the dotted line means a single photon inserted in the cavity and \\
all curves mean external laser fields subjected to atoms}
\end{center}
\vspace{5mm}

It seems to us that applying some results in the paper to the model is 
not difficult. This is also our forthcoming target.

\vspace{10mm}
\noindent{\em Acknowledgment.}\\
We wishes to thank Mikio Nakahara for his helpful comments and suggestions.



\begin{thebibliography}{99}
%
\bibitem{YS}T. Yamanouchi and M. Sugiura : 
\newblock Introduction to Topological Groups (in Japanese), 
\newblock Baifukan, 1960.
%
\bibitem{YMa}Y. Makhlin : 
\newblock Nonlocal properties of two--qubit gates and mixed states 
and optimization of quantum computations, 
\newblock Quant. Info. Proc. 1 (2002), 243, 
\newblock quant-ph/0002045.
%
\bibitem{ZVWS}J. Zhang, J. Vala, K. B. Whaley and S. Sastry :
\newblock A geometric theory of non--local two--qubit operations, 
\newblock Phys. Rev. A 67 (2003), 042313, 
\newblock quant-ph/0209120.
%
\bibitem{RZ}V. Ramakrishna and H. Zhou : 
\newblock On The Exponential of Matrices in $su(4)$, 
\newblock math-ph/0508018.
%
\bibitem{FOS}K. Fujii, H. Oike and T. Suzuki :
\newblock Universal Yang--Mills Action on Four Dimensional Manifolds, 
\newblock Int. J. Geom. Methods Mod. Phys, vol.3, no.7 (2006), 1331, 
\newblock quant-ph/0602204.
%
\bibitem{YM}C. N. Yang and R. L. Mills :
\newblock Conservation of isotopic spin and isotopic gauge invariance, 
\newblock Phys. Rev. 96 (1954), 191.
%
\bibitem{BI}M. Born and L. Infeld :
\newblock Foundations of the new field theory, 
\newblock Proc. Royal Soc. (London), A 144 (1934), 425.
%
\bibitem{GT}V. N. Gorbachev and A. I. Trubilko : 
\newblock Quantum teleportation of EPR pair by three--particle entanglement, 
\newblock J. Exp. Theor. Phys. 91 (2000) 894, 
\newblock quant-ph/9906110.
%
\bibitem{KF3}K. Fujii : 
\newblock Introduction to Coherent States and Quantum Information Theory, 
\newblock quant-ph/0112090.
%
\bibitem{FHKSW}K. Fujii, K. Higashida, R. Kato, T. Suzuki and Y. Wada : 
\newblock Explicit Form of the Evolution Operator of Tavis--Cummings Model : 
Three and Four Atoms Cases, 
\newblock Int. J. Geom. Methods Mod. Phys, vol.1, no.6 (2004), 721, 
\newblock quant-ph/0409068. 
%
\bibitem{KF2}K. Fujii : 
\newblock Introduction to Grassmann Manifolds and Quantum Computation, 
\newblock J. Applied Math, 2 (2002), 371, 
\newblock quant-ph/0103011. 
%
\bibitem{MN}M. Nakahara : 
\newblock Geometry, Topology and Physics, 
\newblock IOP Publishing Ltd, 1990.
%
\bibitem{FHKW1}K. Fujii, K. Higashida, R. Kato and Y. Wada : 
\newblock Cavity QED and Quantum Computation in the Weak Coupling Regime, 
\newblock J. Opt. B: Quantum and Semiclass. Opt, 6 (2004) 502, 
\newblock quant-ph/0407014. 
%
\bibitem{FHKW2}K. Fujii, K. Higashida, R. Kato and Y. Wada : 
\newblock Cavity QED and Quantum Computation in the Weak Coupling Regime II : 
Complete Construction of the Controlled--Controlled NOT Gate, 
\newblock in the book ``Trends in Quantum Computing Research", Chapter 8, 
2006, Nova Science Publishers, Inc (USA), 
\newblock quant-ph/0501046. 
%
\end{thebibliography}
\end{document}